\documentclass[doublecol,figures]{epl2}
\usepackage{amsmath}
\usepackage{latexsym}
\usepackage{epstopdf}
\usepackage{amsmath}
\usepackage{amssymb,amsmath}
\usepackage{multirow}
\usepackage{color}
\newcommand{\beq}{\begin{equation}}
\newcommand{\eeq}{\end{equation}}

\title{Entanglement entropy in long-range harmonic oscillators}
\shorttitle{Entanglement entropy in long-range harmonic oscillators} 

\author{ M. ~Ghasemi ~Nezhadhaghighi\inst{1} \and M.~A.~Rajabpour\inst{2,3}}
\shortauthor{M. ~G. ~Nezhadhaghighi and M.~A.~Rajabpour}

\institute{                    
  \inst{1} Department of Physics, Sharif University of Technology, Tehran,  11365-9161, Iran\\
  \inst{2} SISSA and INFN, \textit{Sezione di Trieste},  via Bonomea 265, 34136 Trieste, Italy \\
  \inst{3} Instituto de F\'{\i}sica de S\~{a}o Carlos, Universidade de S\~{a}o Paulo,
Caixa Postal 369, 13560-970 S\~{a}o Carlos, SP, Brazil
}

\abstract{
We study the Von Neumann and R\'enyi  entanglement entropy of long-range  harmonic oscillators (LRHO) by both theoretical and numerical means. We show that the entanglement entropy in massless harmonic oscillators  increases logarithmically with the sub-system size as $S=\frac{c_{eff}}{3}\log l$. Although the entanglement entropy of LRHO's shares some similarities with the entanglement entropy at conformal critical points we show that the R\'enyi  entanglement entropy presents some deviations from the expected conformal behaviour. In the massive case we demonstrate that the   behaviour of the entanglement entropy with respect to the correlation length is also logarithmic as the short range case.
}

\begin{document}

\maketitle
\section{Introduction}

While the entanglement entropy of short-range interacting quantum systems in one dimension is
well studied  with many different techniques \cite{reviews}, such as  hamiltonian techniques
\cite{Bombelli,Srednicki, Vidal, Peschel}, euclidean methods \cite{Callan,Casini} and conformal
field theory \cite{Holzhey,Cardy1}   there are few results concerning the long-range interacting systems.
The main difficulty is the lack of exact solution for these systems which makes the problem much
 harder than the short-range counterparts. Because of the presence of non-trivial dynamical exponents
  the common euclidean techniques are usually useless in calculating the entanglement entropy in these
  systems. It is worth mentioning that although having a non-trivial dynamical exponents is not restricted
   to just long-range interacting systems, see for example \cite{Refael,Doyon}, they usually provide a
   very natural knob to change it arbitrarily.
Because of the huge finite size effects in these systems even the numerical calculations are  not easy as
their short-range counterparts. Nevertheless, recently much progress has been made in different directions:
 In \cite{LORV} the entanglement entropy in Lipkin-Meshkov-Glick (LMG) model is studied. This model has hamiltonian
  similar to the $XY$ model but while in the latter model the interaction only takes
place between nearest neighbors, in the LMG model, all spins interact among themselves.
In \cite{DHHLB}, the interactions are restricted to the Ising-type without external magnetic field
 which allows  to study both the static and the dynamical entanglement properties of the system. Eisert \textit{et. all} \cite{GESL} found  a logarithmically divergent geometric entropy in free
      fermions with long-range unshielded Coulomb interaction. Plenio \textit{et.all} \cite{Plenio1}
      studied the general properties of the entanglement entropy for interacting harmonic oscillators.
       In particular they found that the area law should be valid in higher dimensions even for long-range
       interacting harmonic oscillators, and most recently the entanglement entropy for the long range
        anti-ferromagnetic Ising chain is calculated by numerical means
        \cite{Tagliacozzo}. {Finally we should mention that in
        \cite{cirac} based on the matrix product states it was
        argued that for those long-range systems that do not have
        any short-range counterparts, in other words one can not
        approximate the ground state of the long-range model with
        the ground state of another short-range model, the presence
        of the long-range interaction implies larger entanglement
        entropy or the volume scaling of the entropy.}

\begin{figure} [htbp]
\centering
\includegraphics[width=8.6cm]{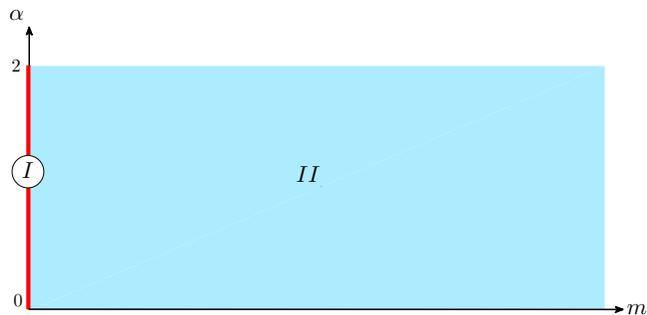}
{\caption{(colour online).  Phase diagram of LRHO in the parameter
space $(\alpha,m)$. The gapless region $(I)$ is highlighted by bold
red line. The system in the region $(II)$ is gapped.
 }}
\label{Figure:1}
\end{figure}

 Harmonic oscillators are the building blocks of many quantum mechanical and field theoretical systems and of course this is also true  for most of the long-range interacting quantum systems and non-local field theories. In this work we study the entanglement entropy of long-range harmonic oscillators in the massless and massive cases.
The main approach we have used is the so called  hamiltonian technique, which was first introduced in \cite{Bombelli} and then elaborated in \cite{Callan}, for review see \cite{Peschel,Casini}. The important advantage of this method with respect to the others in calculating the entanglement entropy in our problem is that, firstly it is possible, to a large extent, to carry out the calculations exactly, and on top of that the numerical calculations are very easy to implement. We will consider the lattice and continuum calculations in parallel. We first introduce the method and then apply it directly to our problem. In the analytical side we will write the main eigenvalue problem exactly, then we will try to find approximate solutions by using our numerical calculations. Using the solutions of the eigenvalue problem we then also calculate the R\'enyi  entanglement entropy and finally we will solve numerically the problem of the massive interacting long-range harmonic oscillators.

\section{Results}

The entanglement entropy of a subsystem $A$ is defined by using reduced density matrix $\rho_A$ as
\begin{equation} \label{EE}
S_A=- \text{tr}\rho_A\log\rho_A.
\end{equation}
In this work we consider three main kinds of subsystems. In the massless case we consider two kinds of systems: in the first one the system is large and  $A$ is a small sub-system with length $l$ and in the second case the system is finite with length $L$ and the sub-system is just half  of it with length $L/2$. In the massive case the system is large and the sub-system is just half of the system. We will provide approximate formula just for the first case and study the two other cases  numerically.

We define the hamiltonian of  one dimensional interacting harmonic oscillators as

\begin{equation} \label{long-rang hamiltonian}
{H=\frac{1}{2}\sum_{i=1}^{N}\dot{\phi_i}^2+\frac{1}{2} \sum_{i,j=1}^{N}\phi_{i} K_{ij}\phi_{j}},
\end{equation}
where  on the lattice we take
\begin{eqnarray} \label{long-rang K}
\begin{split}
K_{i,j} &= {\int_0^{2\pi} \frac{dq}{2\pi} e^{iq(i-j)} \lbrace -\left[ 2(1-\cos(q))\right]^{\frac{\alpha}{2}}+m^{\alpha}\rbrace}\\
&=\frac{\Gamma(-\frac{\alpha}{2}+i-j)\Gamma(\alpha +1)}{\pi \Gamma(1+\frac{\alpha}{2}+i-j)} \sin(\frac{\alpha}{2}\pi)+{m^{\alpha}\delta_{i,j}},
\end{split}
\end{eqnarray}

\begin{figure} [htbp]
\centering
\includegraphics[width=8.6cm]{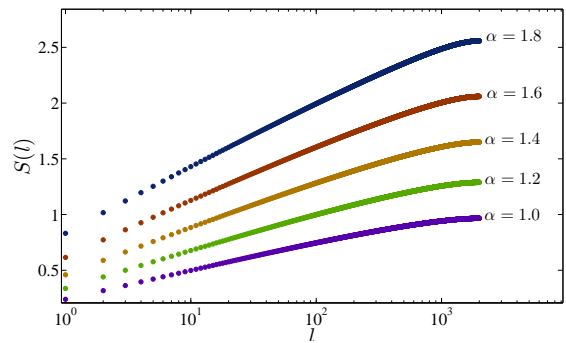}
\caption{(colour online). Entanglement entropy for 1D LRHO: entanglement entropy changes linearly  in the semilog plot in the region $0<l<L/100$, here we have $L=4000$. From top to bottom $\alpha=1.8, 1.6, 1.4, 1.2, 1.0$.}
\label{Figure:2}
\end{figure}

In the special case when $\alpha/2$ is an integer, $K(n) = (-1)^{\alpha-n+1}C_{\alpha,\frac{\alpha}{2}+n}$,
where $C_{\alpha,\frac{\alpha}{2}+n}$ are binomial coefficients. In this
case we remark that $K(n)=0$ for $n>\alpha/2$. It is easy to see that
 in the special case $\alpha=2$  the $K$ matrix is just a simple laplacian. For
 non-integer values for large distances we have $K(n)\sim \frac{1}{n^{1+\alpha}}$.

{In Fig~1 we depicted the phase diagram of the model which is gapless
for $m=0$ and gapped for non-zero value of $m$ for any value of
$\alpha$.}

In the continuum limit  the above hamiltonian can be written as
\begin{equation} \label{fractional free field theory}
\frac{1}{2} \sum_{i,j=1}^{N}\phi_{i} K_{ij}\phi_{j}\rightarrow \int [-\frac{1}{2}\phi(x)(-\bigtriangledown)^{\alpha/2}\phi(x)+\frac{1}{2}m^{\alpha}\phi^2]dx,
\end{equation}
where $-(-\bigtriangledown)^{\alpha/2}$ is the fractional laplacian defined by its Fourier transform $|q|^{\alpha}$.

To measure the entanglement entropy of a sub-system with length $l$ of an infinite system we follow the method explained in \cite{Callan}. First we make the matrices $W=K^{1/2}$ and $W^{-1}=K^{-1/2}$. In the continuum limit they have the following forms
\begin{equation} \label{W matrices}
\begin{split}
 W^{\pm 1}(x,y)&=\frac{1}{2\pi}{\int_{-\infty}^{\infty}dk} (|k|^{\alpha}+m^{\alpha})^{\pm 1/2}e^{ik.(x-y)}\\
 &=\frac{1}{2\Gamma[\mp \alpha/2]\cos(\frac{\pi\alpha}{4})}\frac{1}{(x-y)^{1\pm \alpha/2}}+\mathcal{O}(m^{\alpha}).
\end{split}
\end{equation}

\begin{figure} [htbp]
\centering
\includegraphics[width=8.6cm]{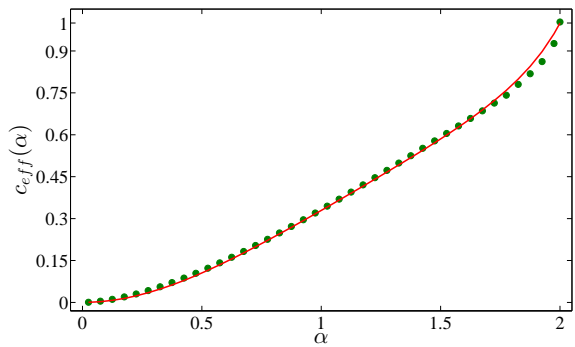}
{\caption{(colour online). Green (Dark gray) circles represent $c_{eff}(\alpha)$ for discrete systems with size $L=6000$. The prefactor $c_{eff}(\alpha)$ is measured using the scaling relation $S$ with $\log l$ in the region $0<l<L/100$. The red line represents the same quantity coming from the continuum limit approximation.}}
\label{Figure:3}
\end{figure}

Then we define the matrix $\Lambda$ by multiplying $W$ and $W^{-1}$ {in the complement of the sub-region $l$}
\begin{equation} \label{Lambda}
\begin{split}
 \Lambda(x,y)&=-\int_{\Omega}dz\Big{(} W^{-1}(x,z)W(z,y)\Big{)}\\
 &=\mathcal{A}\left[\frac{2 \left(\left(\frac{l-x}{l-y}\right)^{\alpha/2}-\left(\frac{x}{y}\right)^{\alpha/2}\right)}{\alpha (x-y)}\right],
 \end{split}
\end{equation}
where {$\Omega\in\left(-\infty<z<0\right)\cup \left( l<z<\infty\right)$ and}   $\mathcal{A}=\frac{1}{4\Gamma[-\alpha/2]\Gamma[\alpha/2]\cos^2(\frac{\pi\alpha}{4})}$. For $\alpha=2$ the matrix has a different form $\pi^2\Lambda(x,y)=\frac{\left(l-x\right) \log(l-x)-\left(l-y\right) \log(l-y)}{(l-y) (y-x)}-\frac{x \log(x)-y \log(y)}{(y-x) y}$. We campaired the matrix $\Lambda(x,y)$ coming from the above equation with numerical $\Lambda$ and  found a very good agreement when the distances are more than four lattice sizes. The agreement gets better by increasing the size of the system.

The entanglement entropy can be expressed directly in terms of the eigenvalues $E_i$ of the matrix $\Lambda$ as \cite{Bombelli,Callan}
\begin{equation} \label{entropy}
S=\sum_{E_i} [\log \frac{\sqrt{E_i}}{2}+\sqrt{1+E_i} \log(\frac{1}{\sqrt{E_i}}+\sqrt{1+\frac{1}{E_i}})].
\end{equation}
In the continuum limit one needs to solve the following eigenvalue problem
\begin{equation} \label{eigenvalue problem}
\int dy \Lambda(x,y)\psi(y)=E \psi(x).
\end{equation}

At the numerical level we followed the above algorithm and calculated the entanglement entropy for large enough system sizes such as $L=5000$ and $L=6000$, to avoid any finite size effects \cite{GR}. For these sizes the data show stability even for very small $\alpha$'s as far as we take the sub-system size less than $L/100$. Interestingly the entanglement entropy grows logarithmically with the sub-system size as
\begin{equation} \label{c_{eff}}
S=\frac{c_{eff}}{3}\log l,
\end{equation}

\begin{figure} [htbp]
\centering

\includegraphics[width=8.6cm]{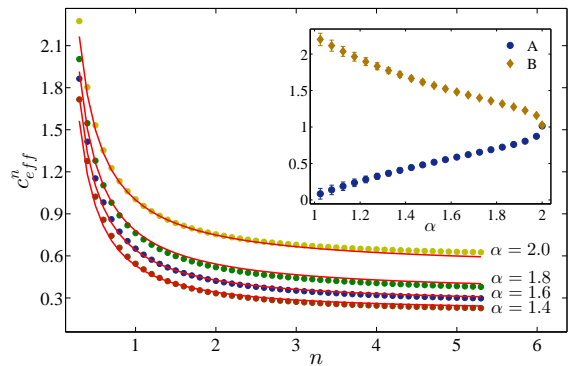}

{\caption{(colour online). $c_{eff}^n$ versus $n$ for different $\alpha$'s (from top to bottom $\alpha=2.0, 1.8, 1.6, 1.4$.). The prefactor  $c^n_{eff}(\alpha)$ is measured using the scaling relation of $S_n$ with $\log l$ in the region $0<l<L/100$ (The system size $L=6000$). Red solid lines come from the continuum limit approximation using the equation (\ref{eigenvalue}). Inset: $A$ and $B$ coefficients versus $\alpha$.}}
\label{Figure:4}
\end{figure}

where $c_{eff}$ is $\alpha$ dependent, see Fig~2. The $c_{eff}$  is
maximum at the conformal short-range point and it starts to decrease
by decreasing $\alpha$.

For $\alpha=2$ Callan and Wilczek \cite{Callan} found a good
estimation for the
 eigenvalues of the $\Lambda$ operator when the subsystem is half of a finite system.
 Motivated by their work we found $E(\omega)=\frac{1}{\sinh^2(\pi\omega)}$, where $\omega(E_i)\log( l)=\frac{\pi i}{2}$
  a very good approximation for the eigenvalues of the short-range case.
   Based on numerical comparison we found that apart from a constant the  behaviour
   of the  logarithm of the small eigenvalues (large $i$, i.e. $i>6$) is independent
   of $\alpha$ and one can safely conjecture the following behaviour for the eigenvalues of the $\Lambda$ operator

\begin{equation} \label{eigenvalue}
E(\omega)=\frac{a(\alpha)}{\sinh^2(\pi\omega)+b(\alpha)},
\end{equation}
where $a(\alpha)$ and $b(\alpha)$ are functions of $\alpha$ and $\omega\log( l)=\frac{\pi i}{2}$. Non-zero value of $b(\alpha)$ is necessary to make the formula compatible with larger eigenvalues. We found $a(\alpha)=\frac{\alpha}{2}\sin^2(\frac{\pi\alpha}{4})$   and $b(\alpha)= 0.1146\alpha+0.1868\alpha^2-0.1988\alpha ^3+0.03771\alpha ^4
$  best fit functions to our numerical data. The value of $b(\alpha)$ is zero at $\alpha=0$ and $2$ and it has a maximum at $\alpha=1$. In principle $b(\alpha)$ might be not a simple polynomial.
Using the above formula one can get the logarithmic behaviour for the entanglement entropy for free and  the $c_{eff}$ is
\begin{equation} \label{c_{eff}}
\begin{split}
 c_{eff}(\alpha)&=\frac{6}{\pi}\int_{0}^{\infty}d\omega \Big{[}\log\frac{\sqrt{E(\omega)}}{2}
 +\sqrt{1+E(\omega)}\\
 &\times\log(\frac{1}{\sqrt{E(\omega)}}
 +\sqrt{1+\frac{1}{E(\omega)}})\Big{]}.
\end{split}
\end{equation}

\begin{figure} [htbp]
\centering
\includegraphics[width=8.6cm]{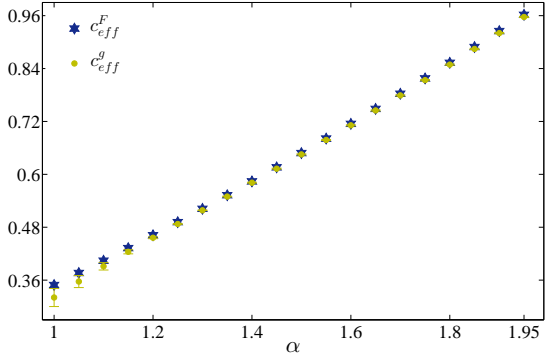}
{\caption{(colour online).  The dark blue (black) stars are the prefactors of the entanglement entropy for the massless system with length $L$, i.e. $c^{F}_{eff}=6S/\log L$, ($L\in\left( 1000-4000 \right)$). The light green (light gray) circles are the prefactors for the entanglement entropy of the massive systems, i.e. $c^{g}_{eff}=6S/\log m$. The quantity $c^{g}_{eff}$ is measured using the best fit to $S$ in the saturation regime versus $\log m$, where $m\in\left( 10^{-3}-10^{-2} \right)$ and the system size $L=4000$.}}
\label{Figure:5}
\end{figure}

In Fig ~3 we compare the result coming from the above formula and the numerical calculations, with  excellent agreement. The logarithmic behaviour of the entanglement entropy is  reminiscent of the presence of conformal symmetry. To check this point we calculated the entanglement R\'enyi  entropy using the same technique. The
R\'enyi  entropy $S_n$ is defined as

\begin{equation} \label{RE}
S_n = \frac{1}{1-n}\log tr \rho^n, \hspace{1cm} n>1
\end{equation}
and can be calculated using the folowing formula \cite{Casini}
\begin{equation}\label{GREntropy}
{S_n = \frac{1}{n-1}\sum_i \left( \log(1-\xi^n) -n\log(1-\xi) \right)}.
 \end{equation}
where {$\xi=\frac{\sqrt{1+E_i}-1}{\sqrt{1+E_i}+1}$} with $E_i$ as
the eigenvalues of the  $\Lambda$ matrix. In conformal invariant
systems $S$ changes logarithmically with a prefactor
$\frac{c}{6}(1+\frac{1}{n})$, where $c$ is the central charge
\cite{Cardy1}. We found that in LRHO the R\'enyi  entropy is also
increases with the sub-system size logarithmically as
$S_n=\frac{c_{eff}^n}{3}\log(l)$. We  have depicted $c_{eff}^n$
versus $n$ in Fig~4. The best fit is
$c_{eff}^n=\frac{c_{eff}}{2}(A(\alpha)+B(\alpha)/n)$ where
$A(\alpha)$ and $B(\alpha)$ are functions of $\alpha$. The $\alpha$
dependency of these coefficients shows that this system is not
conformally invariant except at $\alpha=2$ where we have $A=B=1$.
{This is not surprising because it is well-known that  in long-range
systems we expect conformal symmetry just when the dynamical
critical exponent $z$ is equal to one while in our system the
dynamical critical exponent is $z=\frac{\alpha}{2}$ \cite{dutta}.}

In the continuum limit one can write (\ref{GREntropy}) as
\begin{equation}\label{GREntropy2}
 S_n = \frac{{\log(l)}}{\pi(n-1)}\int_0^\infty d\omega  \left( \log(1-\xi^{n}) -n\log(1-\xi) \right),
 \end{equation}
where $\xi=\frac{\sqrt{1+E(\omega)}-1}{\sqrt{1+E(\omega)}+1}$ and $E(\omega)$ comes from Eq. (\ref{eigenvalue}). After doing
the integration numerically one can find $c_{eff}^n$ versus $n$, see Fig ~4. The perfect agreement between numerical
calculation and the continuum limit results is another support for the accuracy of the Eq. (\ref{eigenvalue}).

{Now we discuss the effect of boundary condition to the
entanglement entropy.} The most interesting one is the free boundary
condition, which we take a finite system with half of it as the subsystem.
In this case we define the $K$ matrix by throwing away the elements of the
 infinite matrix which is not inside the system. The immediate problem that
  we face is then that we are not able to write the $W$ matrices analytically
   because in the presence of the boundary we are not allowed to use Fourier transform.
   In this case we follow all the calculations numerically. The algorithm is the same
    and interestingly we found again logarithmic behaviour for the entanglement entropy
    with respect to the size of the system $\frac{c^{F}_{eff}}{6}\log L$.
    The prefactor $c^{F}_{eff}$ is in general different from $c_{eff}$ except
    at conformal point, see Fig~5. This fact indicates that in the long-range
    interacting systems the entanglement entropy of a sub-system is not equal to
    the summation of the contribution of  each boundary of the sub-system with its complement.

Finally we report the entanglement entropy of massive long-range interacting harmonic oscillators.
The entanglement entropy  of the positive half of an infinite massive system  for short-range models comes
from the Cardy- Calabrese formula $S=-\frac{c}{6}\log m$ where $c$ is the central charge of the system and
it is one for short-range harmonic oscillators \cite{Cardy1}. We calculated the same quantity for the long-range
interacting systems. We found that $S$ saturates in $l\rightarrow \infty$  limit, and changes logarithmically with respect to the mass as
\begin{equation}\label{GREntropy2}
 S=-\frac{c^{g}_{eff}}{6}\log m,
 \end{equation}
where $c^{g}_{eff}$ is different from $c_{eff}$ except at the
$\alpha=2$. {We should mention that the presence of the pseudo
central charge in long-range systems was already reported in
\cite{Tagliacozzo}. It is worth to mention that interestingly
$c^{g}_{eff}$ is equal to the prefactor of the free boundary
condition case, see Fig ~5. This can be understood as follows: In
the definition of the $K$ matrix for free boundary condition we
assumed that we just throw away those elements of the infinite
system which are not inside the corresponding finite system which
makes the summation of the every row of the matrix non-zero. This
can show itself as an effective mass. One can check this guess by
looking to the entanglement entropy of a subsystem of a large but
finite system. By throwing away some elements of the infinite $K$
matrix we make the system gapped then one can easily check that one
can introduce the same amount of gap by putting some mass in the
infinite system. The corresponding mass is equivalent to correlation
length $\zeta=\frac{1}{m^{\alpha/2}}$. As far as one takes a
subsystem smaller than this length one will get just $c_{eff}$ but
if we take the subsystem bigger than this length then the boundary
effect will show itself as the mass gap.}

 Finally it is worth
mentioning that since in the long-range interacting systems the
correlation length
 changes as {$\frac{1}{m^{\alpha/2}}$} one might expect that the the
entanglement
 entropy be proportional to $\log \zeta$ and then expect that $c^{g}_{eff}$ be proportional to $\alpha$. Although it seems that
  this is the case in the region $1<\alpha<2$ the slop that we find is not compatible with the natural expectation $c^{g}_{eff}=0$
  at $\alpha=0$. Moreover our primary numerical calculations  show strong deviation from the linear behaviour close to $\alpha=0$

\section{Conclusion and outline}:
In this paper we have found the entanglement entropy of long-range
harmonic oscillators using the hamiltonian technique. When we
consider a sub-system with length $l$ the entanglement entropy
follows the logarithmic behaviour as the short range cases but the
prefactor is dependent on the power of the interaction $\alpha$.

 Although it is tempting to interpret the prefactor of
the logarithm $c_{eff}$ as the effective central charge of the system,
by calculating the R\'enyi entropy we showed that the nature of this prefactor
is  different from the central charge at least at the level of the R\'enyi entropy.

{
This result is  in contradiction with the
theorem proved in \cite{cirac}, the reason is that the ground state of our long-range hamiltonian does not necessarily related to the ground state of any short-range hamiltonian! This is simply because if it was possible to approximate the ground state of our hamiltonian with  ground state of a short-range hamiltonian we would expect conformal symmetry which is not the case in our model. The nature of this discrepancy is not clear for us.
}

We also calculated entanglement entropy for two other interesting cases, the system
 with boundary and also the massive case. Interestingly we found that the prefactor
 is the same in these two different cases, however,  it is different from the massless
  infinite system. Our study shows that the entanglement entropy of long-range interacting
   systems shows some similarities with the entanglement entropy of short-range systems ,
   however, since they have very different symmetries they start to show strong differences
   when we study their replica behaviour. We believe that long-range interacting systems show
    interesting features which deserve more intense studies. Our work can be extended in many
     directions: calculating the entanglement entropy in higher dimensions, the dynamics of the
     entanglement entropy and the finite size effects are just few among many other interesting directions.

\acknowledgments

We are indepted to J. Cardy for many interesting  advices and help. {Thanks are also due to the anonymous referee for many suggestions.} M. G. Nezhadhaghighi thanks S. Rouhani for his supports and also helpful and motivating discussions. M. A. Rajabpour thanks I. Peschel, P. Calabrese, S. Sotiriadis and P. G. de Assis for interesting discussions.  M. A. Rajabpour  thanks the Galileo Galilei Institute for Theoretical Physics for the hospitality and the INFN for partial support during the completion of this work. The work of MAR was supported in part by FAPESP.


\begin{thebibliography}{44}
\bibitem{GR} We discuss the {finite size effects} in this problem in a separate work.
\end{thebibliography}
\end{document}